\documentstyle[preprint,aps,floats]{revtex}
\begin{document}
\input FEYNMAN

\def\tr{\,{\rm tr}}
\def\vp{\varphi}
\def\hphi{\hat{\phi}}
\def\hD{\hat{D}}
\def\hvp{\hat{\varphi}}
\def\pd{{\partial} }

\draft
\preprint{} 
\setcounter{footnote}{1}
\title{The Non-abelian Chern-Simons Coefficient in the Higgs Phase}
\author{Hsien-chung Kao\footnote{Email address:hckao@ibm1.phys.tku.edu.tw}}
\address{Department of Physics, Tamkang University, Tamsui, Taiwan 25137, R.O.C.}

%\date{\today}

\maketitle
\begin{abstract}
We calculate the one loop corrections to the Chern-Simons coefficient $\kappa$ in the Higgs phase of Yang-Mills Chern-Simons Higgs theories.  When the gauge group is $SU(N)$, we show, by taking into account the effect of the would be Chern-Simons term, that the corrections are always integer multiples of ${1\over 4\pi}$, as they should for the theories to be quantum-mechanically consistent.  In particular, the correction is vanishing for $SU(2)$.  The same method can also be applied to the case that the gauge group is $SO(N)$.  The result for $SO(2)$ agrees with that found in the abelian Chern-Simons theories.  Therefore, the calculation provides with us a unified understanding of the quantum correction to the Chern-Simons coefficient.
\end{abstract}
\vfill\eject
%\narrowtext

Chern-Simons theories in 2+1 dimension provide a field-theoretic description of particle excitations with fractional spin and statistics, and thus can be used to study the fractional quantum Hall effect \cite{CSFT,CSANYON,FQHE}.  Furthermore, we can put the theories in a self-dual form by including Higgs fields with a special sixth order potential.  When this happens, the systems admit a so-called Bogomol'nyi bound in energy \cite{Bogo}, which is saturated by solutions satisfying a set of first-order self-duality equations \cite{Hong}.  These solutions have rich structure especially when the gauge symmetry is non-abelian and are interesting in their own right \cite{NASD}.  It is also known that the self-duality in these systems is a result of an underlying $N=2$ supersymmetry \cite{CSII}.

The quantum correction to the Chern-Simons coefficient is also interesting.  When there are neither massless charged particles nor spontaneous symmetry breaking, Coleman and Hill have shown that the correction to the Chern-Simons coefficient can only come from the fermion one-loop effect and is quantized (${1\over 4\pi}$) \cite{Coleman}.  When either of the two conditions is violated, scalar particles may also contribute to the correction and higher-loop effect is generally non-vanishing \cite{Semenoff,Khleb1}.  In particular, the one-loop correction looks complicated and is not quantized.  For abelian Chern-Simons theories, this does not really cause a problem.  

When the gauge symmetry is non-abelian, however, the Chern-Simons coefficient must be integer multiple of ${1\over 4\pi}$ for the systems to be quantum-mechanically consistent.  Therefore, it is interesting to see whether the quantization condition survives the quantum correction.  In the symmetric phase, this has been shown in one-loop \cite{Pisarski}.  In the Higgs phase, the situation is more subtle.  If there is remaining symmetry in the Higgs phase, e.g. $SU(N)$ with $N\ge 3$, it has been shown that the correction still satisfies the quantization condition \cite{Dunne1,Khare,PV}.  In contrast, if the gauge symmetry is completely broken, e.g. $SU(2)$, the correction is again complicated and not quantized \cite{Khleb2}.  It is usually argued that this arises because there is no well-defined symmetry generator in the system.

One can, however, take an alternative perspective about the whole thing.  More careful analysis suggests that there could exist in the effective action so-called would be Chern-Simons terms, which are completely gauge invariant while induce in the Higgs phase terms similar to the Chern-Simons one \cite{Khleb2}.  In fact, the one-loop correction in the Higgs phase has been shown to be identical to that in the symmetric phase for a general class of renormalizable abelian Chern-Simons theories \cite{CHTh}. Although the results reported in Ref. \cite{Dunne2} look complicated, we believe they can also be incorporated in this picture.  Therefore everything will fit together well, if only we could show how it works in the non-abelian theories.  Unfortunately, the calculation seems to be much too complicated, and an explicit demonstration is still lacking.

In this paper, we take up the on-going effort and calculate the one loop corrections to the Chern-Simons coefficient in the Higgs phase of Yang-Mills Chern-Simons Higgs theories.  With the Higgs being in the fundamental $SU(N)$, we show that the corrections are always integer multiples of ${1\over 4\pi}$ for all $N$.  In particular, the correction is vanishing for $SU(2)$.  The nice thing is that we can avoid the tedious calculations encountered in Ref. \cite{CHTh} as will be explained later.  We also apply the same method to the case that the gauge group is $SO(N)$.  In particular, the correction is vanishing for $SO(2)$, consistent with the result in Ref. \cite{CHTh}.  We conclude with some comment on the case when the Higgs field is in the adjoint representation.

Let us consider the following Yang-Mills Chern-Simons theories with a complex Higgs field $\Phi$ in the fundamental representation:
\begin{eqnarray} 
{\cal L} = {1\over g^2}\tr \biggl\{ 
&\;& -{1 \over 2g^2}F^2_{\mu \nu} - i\kappa \; \epsilon^{\mu\nu\rho}
(A_\mu \partial_\nu A_\rho - {2\over 3}iA_\mu A_\nu A_\rho) \biggr\} \nonumber \\
&\;& + |D_\mu \Phi|^2 + \lambda (|\Phi|^2- v^2)^2. 
\label{Lg}
\end{eqnarray}
Here $D_\mu = (\partial_\mu - i A^m_\mu T^m)$ and $\epsilon_{012} = 1$.  To be specific, we choose the gauge group to be $SU(N)$.  The generators satisfy $[T^m, T^n] = if^{lmn}T^l,$ with the normalization $\tr\{T^m T^n\} = \delta^{mn}/2$.  Moreover, $\sum_m (T^m)_{\alpha\beta}(T^m)_{\gamma\delta} ={1\over 2}\delta_{\alpha\delta} \delta_{\beta\gamma} - {1\over 2N}\delta_{\alpha\beta} \delta_{\gamma\delta}.$ 

Because of it conceptual advantage, the background field method will be employed.  For this purpose, we separate $A_\mu$ into the background part $A_\mu$ and the quantum part $Q_\mu.$  In the Higgs phase, $\Phi = \phi + \vp$ with $\vp^\dagger\vp = v^2$.  The gauge fixing term is given by
\begin{equation}
{\cal L}_{gf}={1\over 2\xi}\biggl\{\bigl(\hD_\mu Q_\mu\bigr)^m
+ i\xi\bigl(\vp^\dagger T^m\phi - \phi^\dagger T^m\vp \bigr) \biggr\}^2,
\label{Lgf}
\end{equation}
where $\hD_\mu$ is the covariant derivative with the background field.
Following standard procedure, one can find the Faddeev-Popov ghost term:
\begin{eqnarray}
{\cal L}_{FP}  = 
&\;& 2\,\tr \biggl\{\bigl(\hD_\mu \bar{\eta} \bigr) \bigl(\hD_\mu \eta\bigr)
-i\bigl(\hD_{\mu} \bar{\eta} \bigr) \bigl[Q_{\mu}, \eta \bigr] \biggr\}\nonumber\\
&\;& + \xi\bigl(\vp^\dagger\bar{\eta}\eta\vp - \vp^\dagger\eta\bar{\eta}\vp\bigr)
+ \xi\bigl(\vp^\dagger\bar{\eta}\eta\phi - \phi^\dagger\eta\bar{\eta}\vp\bigr)
\label{Lgh}
\end{eqnarray}
From Eqs. (\ref{Lg}), (\ref{Lgf}) and (\ref{Lgh}), we see to quadratic order in $Q_\mu$ and $\phi,$ the revelant terms are
\begin{eqnarray} 
{\cal L}_0 = 
&\;& {1\over 2} Q^m_\mu \biggl\{
\bigl[{-1\over g^2}(\pd^2\delta_{\mu\nu} - \pd_\mu \pd_\nu) 
- {1\over \xi}\pd_\mu \pd_\nu 
+ i\kappa \epsilon_{\mu\nu\rho} \pd_\rho \bigr] \delta_{mn} \nonumber\\
&\;& \qquad \qquad \qquad \qquad \qquad \qquad + \delta_{\mu\nu}\bigl[
(\vp^\dagger T^mT^n\vp) + (\vp^\dagger T^nT^m\vp)\bigr]\biggr\}Q^n_\nu \nonumber\\
&\;& + {1\over 2} \phi^p_a \biggl\{
\bigl[-\pd^2 + {1\over 2} \xi \vp^2  \bigr]
\bigl( \delta_{ab}\delta_{pq}- \hvp^p_a \hvp^q_b \bigr) 
+ \bigl[-\pd^2 + m_H^2 \bigr] \bigl(\hvp^p_a \hvp^q_b \bigr) \nonumber\\
&\;& \qquad \qquad \qquad \qquad \qquad \qquad \qquad \qquad 
+ \bigl[{(N-2) \over 2N}\xi\vp^2\bigr] 
\bigl(\epsilon_{ac}\epsilon_{bd} \hvp^p_c \hvp^q_d \bigr) \biggr\} \phi^q_b \label{LBq}\\
&\;& + f^{lmn} \biggl\{{1\over g^2}(\pd_\mu A^l_\nu) Q^m_\mu Q^n_\nu
+ {1\over g^2}(\pd_\mu Q^l_\nu) A^m_\mu Q^n_\nu
+ {1\over g^2}(\pd_\mu Q^l_\nu) Q^m_\mu A^n_\nu \nonumber\\
&\;& \qquad \qquad + {1\over \xi}(\pd_\mu Q^l_\mu) A^m_\mu Q^n_\nu
- {i\kappa\over 2} \epsilon_{\mu\nu\rho} A^l_\mu Q^m_\nu Q^n_\rho \biggr\} \nonumber\\
&\;& + 2 (\vp^\dagger A_\mu Q_\mu \phi) + 2 (\phi^\dagger Q_\mu A_\mu \vp). \nonumber
\end{eqnarray}
Here, $\phi^p = {1\over \sqrt{2}}(\phi^p_1 + i \phi^p_2),$ 
$\vp^p = (\vp^p_1 + i \vp^p_2),$ $\hvp^p_a = \vp^p_a/\sqrt{\vp^2}$, with $p,q = 1,2,\ldots, N$ denoting the components of the Higgs field.  $m_H^2= 4\lambda\vp^2$, with $\vp^2 = \sum_{p,a}(\vp^p_a)^2 = v^2.$

As pointed out in \cite{Khleb2}, there could exist in the effective action the so-called would be Chern-Simons terms, which are invariant even under the large gauge transformation while induce terms similar to the Chern-Simons one in the Higgs phase.  In fact, one finds there are two such terms revelant to our discussion:
\begin{eqnarray}
&\;& O_1 = \epsilon^{\mu\nu\rho} 
i\bigl\{\Phi^\dagger T^m (D_\mu \Phi) -(D_\mu\Phi)^\dagger T^m \Phi \bigr\}
F^m_{\nu\rho}, \nonumber\\
&\;& O_2 = \epsilon^{\mu\nu\rho} 
i\bigl\{\Phi^\dagger (D_\mu \Phi) - (D_\mu\Phi)^\dagger \Phi \bigr\}
(\Phi^\dagger F_{\nu\rho} \Phi). 
\end{eqnarray}
In the Higgs phase, they give rise to  
\begin{eqnarray}
&\;& \epsilon^{\mu\nu\rho} A^n_\mu F^m_{\nu\rho}
\bigl\{(\vp^\dagger T^m T^n\vp) + (\vp^\dagger T^n T^m\vp)\bigr\},
\nonumber\\
&\;& 2\epsilon^{\mu\nu\rho} A^n_\mu F^m_{\nu\rho}
(\vp^\dagger T^m\vp)(\vp^\dagger T^n \vp), 
\label{WBCS}
\end{eqnarray}
respectively.

To extract the correction to the Chern-Simons coefficient, it is helpful to introduce the following projection operators in finding the propagators of the gauge and Higgs fields:
\begin{eqnarray} 
&\;& (P_1)_{mn} = \delta_{mn} 
- 2 \bigl[(\hvp^\dagger T^mT^n \hvp) + (\hvp^\dagger T^nT^m \hvp) \bigr]
+ {2(N-2)\over (N-1)} (\hvp^\dagger T^m \hvp) (\hvp^\dagger T^n \hvp), \nonumber\\
&\;& (P_2)_{mn} =  
2 \bigl[(\hvp^\dagger T^mT^n \hvp) + (\hvp^\dagger T^nT^m \hvp) \bigr]
- 4 (\hvp^\dagger T^m \hvp) (\hvp^\dagger T^n \hvp), \nonumber\\
&\;& (P_3)_{mn} = 
{2N \over (N-1)} (\hvp^\dagger T^m \hvp) (\hvp^\dagger T^n \hvp); \nonumber\\
&\;& (R_1)^{pq}_{ab} = \delta_{pq} \delta_{ab} 
- \hvp^p_a \hvp^q_b - \epsilon_{ac} \epsilon_{bd} \hvp^p_c \hvp^q_d, \nonumber\\
&\;& (R_2)^{pq}_{ab} = \hvp^p_a \hvp^q_b \nonumber\\
&\;& (R_3)^{pq}_{ab} = \epsilon_{ac} \epsilon_{bd} \hvp^p_c \hvp^q_d.  
\end{eqnarray}
It is easy to check that they indeed satisfy
\begin{eqnarray} 
&\;& P_i P_j = \delta_{ij}P_i; \nonumber\\
&\;& R_i R_j = \delta_{ij}R_i. 
\end{eqnarray}
With these projection operators, it is now straightforward to obtain the propagators of $Q_\mu$ and $\phi$:
\begin{eqnarray} 
\Delta^{mn}_{\mu\nu}(k) = \biggl\{
&\;& \bigl[\Delta^1_{\mu\nu}(k) \bigr] (P_1)_{mn} 
+ \bigl[\Delta^2_{\mu\nu}(k) \bigr] (P_2)_{mn} 
+ \bigl[\Delta^3_{\mu\nu}(k) \bigr] (P_3)_{mn} \biggr\}, \nonumber\\
D^{pq}_{ab}(k) = \biggl\{
&\;& \bigl[ D^1(k)\bigr] (R_1)^{pq}_{ab} 
+ \bigl[ D^2(k)\bigr] (R_2)^{pq}_{ab} 
+ \bigl[ D^3(k)\bigr] (R_3)^{pq}_{ab} \biggr\}.
\label{Prop}
\end{eqnarray}
Here, 
\begin{eqnarray} 
&\;& \Delta^1_{\mu\nu}(k) = {g^2(k^2 \delta_{\mu\nu} - k_\mu k_\nu) 
+ g^2 M\epsilon_{\mu\nu\rho} k^\rho \over k^2(k^2 + M^2)} 
+ {\xi k_\mu k_\nu\over k^4}, \nonumber\\
&\;& \Delta^2_{\mu\nu}(k) = {g^2(k^2 + M_{W^+} M_{W^-}) 
(\delta_{\mu\nu} - k_\mu k_\nu/k^2) + g^2M\epsilon_{\mu\nu\rho} k^\rho 
\over (k^2 + M_{W^+}^2)(k^2 + M_{W^-}^2)} 
+ {\xi k_\mu k_\nu\over k^2(k^2 + {1\over 2} \xi \vp^2) }, \nonumber\\
&\;& \Delta^3_{\mu\nu}(k) = {g^2(k^2 + M_{Z^+} M_{Z^-}) 
(\delta_{\mu\nu} - k_\mu k_\nu/k^2) + g^2M\epsilon_{\mu\nu\rho} k^\rho 
\over (k^2 + M_{Z^+}^2)(k^2 + M_{Z^-}^2)} 
+ {\xi k_\mu k_\nu\over k^2\bigl[k^2 + {(N-1)\over N} \xi \vp^2\bigr] }, \nonumber\\
&\;& D^1(k) =  {1\over (k^2 + {1\over 2} \xi \vp^2)}, \nonumber\\
&\;& D^2(k) = {1 \over (k^2 + m_H^2)}, \nonumber\\
&\;& D^3(k) =  {1 \over \bigl[k^2 + {(N-1) \over N} \xi \vp^2 \bigr]}, 
\end{eqnarray}
with $M= \kappa g^2,$ and 
\begin{eqnarray}
&\;& M_{W^\pm} = (a_W \pm 1)M/2, \quad a_W = \sqrt{1 + {2\vp^2\over \kappa^2 g^2}}; \nonumber\\
&\;& M_{Z^\pm} = (a_Z \pm 1)M/2, \quad a_Z = \sqrt{1 + {{4(N-1)\over N}\vp^2\over \kappa^2 g^2}}. 
\end{eqnarray}
Note that $\Delta^1_{\mu\nu},\Delta^2_{\mu\nu}$ and $\Delta^3_{\mu\nu}$ correspond to propagators of the unbroken part, $W$ and $Z$, repectively.

To one-loop order, only the two graphs in Fig. 1 contribute to the parity odd part of the vacuum polarization.  They come from the diagrams with a gauge loop and a gauge-Higgs loop, respectively \cite{PV}.
Carry out the algebra, and we see that
\begin{eqnarray}
\bigl[\Pi^{mn}_{\mu\nu}(p) \bigr]_{odd} = 
\epsilon_{\mu\nu\rho}p_\rho \biggl\{ \Pi_1(p^2) \delta_{mn}
&\;& + \Pi_2(p^2)\bigl[(\hvp^\dagger T^mT^n\hvp)+(\hvp^\dagger T^nT^m\hvp) \bigr]
\nonumber\\
&\;& \qquad \qquad
+ \Pi_3(p^2) (\hvp^\dagger T^m \hvp) (\hvp^\dagger T^n \hvp) \biggr\}. 
\label{PI}
\end{eqnarray}
It is easy to see that the two would be Chern-Simons terms only contribute to $\Pi_2(0)$ and $\Pi_3(0)$.  Therefore, all we need to calculate is $\Pi_1(0)$ to find the correction to the Chern-Simons coefficient.  In the Landau gauge,
\begin{eqnarray}
\Pi_1(p) = &\;& {(N-1)\over 2}\Pi^{Ia}(p) + {1\over 2}\Pi^{Ib}(p), \nonumber\\
\Pi_2(p) = &\;& -(N-1)\Pi^{Ia}(p) - \Pi^{Ib}(p) 
+ {N(N-2)\over (N-1)} \Pi^{Ic}(p) + {N\over (N-1)} \Pi^{Id}(p) \nonumber\\
&\;& + \Pi^{IIa}(p) + \Pi^{IIb}(p) 
+ {2N(N-2)\over (N-1)} \Pi^{IIc}(p) + {2\over N(N-1)} \Pi^{IId}(p), \nonumber\\
\Pi_3(p) = &\;& {(N-2)\over 2}\Pi^{Ia}(p) + {(N+2)\over 2}\Pi^{Ib}(p) 
- {N(N-2)\over (N-1)} \Pi^{Ic}(p) - {N\over (N-1)} \Pi^{Id}(p) \\
&\;& - \Pi^{IIa}(p) - \Pi^{IIb}(p) 
- {2N(N-2)\over (N-1)} \Pi^{IIc}(p) - {2\over N(N-1)} \Pi^{IId}(p) \nonumber\\
&\;& + {2(N-1)\over N} \Pi^{IIe}(p) + 2(N-1) \Pi^{IIf}(p). \nonumber
\end{eqnarray}
Here,
\begin{eqnarray}
\Pi^{Ia}(p) = 
&\;& \int {d^3k\over (2\pi)^3}
\biggl\{ {M\bigl[k^2 p^2 -(k\cdot p)^2 \bigr]
\bigl[ 4M^2 + 10k^2-10k\cdot p + 8p^2 \bigr] \over
p^2 k^2(k^2+M^2)(k-p)^2\bigl[ (k-p)^2+M^2 \bigr]}\biggr\} \nonumber\\
+ &\;& \int {d^3k\over (2\pi)^3} 
\biggl\{ {M\bigl[ -2k^2 p^2 -2(k\cdot p)^2 + 4p^2(k\cdot p) \bigr] \over
p^2 k^2(k^2+M^2)(k-p)^2} \biggr\}, \nonumber\\
\Pi^{Ib}(p) = 
&\;& \int {d^3k\over (2\pi)^3}
\biggl\{ {M\bigl[k^2 p^2 -(k\cdot p)^2 \bigr] \over p^2 
(k^2+M_{W^+}^2)(k^2+M_{W^-}^2)
\bigl[(k-p)^2+M_{W^+}^2\bigr] \bigl[(k-p)^2+M_{W^-}^2\bigr] }\biggr\}\nonumber\\
&\;& \qquad\quad \times \biggl\{ 6M^2 + {(k^2 + M_{W^+} M_{W^-})
\bigl[- M^2 + 8k^2 - 4k\cdot p + 4p^2\bigr]\over k^2} \nonumber\\
&\;& \qquad\qquad + {\bigl[(k-p)^2 + M_{W^+} M_{W^-}\bigr] 
\bigl[-M^2 + 8k^2 - 12k\cdot p + 8p^2 \bigr] \over (k-p)^2 } \label{Pi}\\
&\;& \qquad\qquad + {(k^2 + M_{W^+} M_{W^-})\bigl[(k-p)^2 + M_{W^+} M_{W^-} \bigr]
\bigl[-6k^2 + 6k\cdot p -4p^2 \bigr] \over k^2(k-p)^2 } \biggr\} \nonumber\\
+ &\;& \int {d^3k\over (2\pi)^3} 
\biggl\{ {-2M(k\cdot p)\bigl[M_{W^+} M_{W^-} + 2k^2 - 2p^2 \bigr] \over
p^2 (k^2 + M_{W^+}^2)(k^2+M_{W^-}^2)(k-p)^2} \nonumber\\
&\;& \qquad\qquad + {M(k^2 + M_{W^+} M_{W^-}) 
\bigl[-2k^2p^2 - 2(k\cdot p)^2 + 4k^2(k\cdot p) \bigr]
\over p^2 k^2(k^2 + M_{W^+}^2)(k^2+M_{W^-}^2)(k-p)^2} \biggr\}, \nonumber
\end{eqnarray}
and all other integrals are given in the appendix.  Note that $\Pi^{Ia}$ and $\Pi^{Ib}$ are identical to Eqs. (11) and (12) in Ref. \cite{PV} respetively up to a factor.  Since $\Pi^{Ia}$ and $\Pi^{Ib}$ only involve the diagram with gluon loop, the result we obtain here is actually independent of the form of the Higgs potential.  In the zero momentum limit,
\begin{eqnarray}
&\;& \Pi^{Ia}(0) = {\kappa\over 2\pi|\kappa|}, \nonumber\\
&\;& \Pi^{Ib}(0) = 0. 
\end{eqnarray}

In the background field gauge $Q_\mu$ does not get renormalized.  As a result,
\begin{eqnarray}
\kappa_{\rm ren} 
&\;& = \kappa  + \Pi_1 (0), \nonumber\\
&\;& = \kappa  + {(N-1)\kappa\over 4\pi|\kappa|} 
\end{eqnarray}
for $N\ge 3,$ in agreement with the results found in Ref. \cite{Dunne1,Khare,PV}.  Although the above result can also be obtained by calculating the parity odd part of the vacuum polarization in the unbroken sector as in Ref. \cite{Dunne1,Khare,PV}, this might be particular to the case that the Higgs field is in the fundamental representation.

In the $SU(2)$ case, the gauge symmetry is completely broken and there is no such a thing as unbroken part in the Higgs phase.  As a result, all the terms involving $\Delta^1_{\mu\nu}$ should be set to zero and hence the correction to the Chern-Simons coefficient is vanishing consistent with the claim in \cite{Khare}.  A interesting point is that for $SU(2)$ the group generators $T^m$'s are proportional to the Pauli matrices.  Making use of the identity $(\sigma^m \sigma^n + \sigma^n \sigma^m) = 2\delta_{mn},$ we see that the first term in (\ref{WBCS}) becomes proportional to the Chern-Simons one.  This explains why it is impossible to find the right correction to the Chern-Simons coefficient in the conventional calculation.

We can perform similar calculation in the $SO(N)$ case by noting that there $\tr\{T^m T^n\} = 2\delta^{mn}$, $\sum_m (T^m)_{\alpha\beta}(T^m)_{\gamma\delta} = \delta_{\alpha\delta} \delta_{\beta\gamma} - \delta_{\gamma\alpha} \delta_{\beta\delta},$ and $(\vp^\dagger T^m\vp) =0$.  The procedure is very similar and we will just give the main result: 
\begin{equation}
\Pi_1(p) = (N-3)\Pi^{Ia}(p) + \Pi^{Ib}(p). 
\end{equation}
Here, $\Pi^{Ia}$ and $\Pi^{Ib}$ are identical to those in Eqs.(\ref{Pi}) with $a_W = \sqrt{1 + {8\vp^2\over \kappa^2 g^2}}$.  Consequently,
\begin{equation}
\kappa_{\rm ren} = \kappa  + {(N-3)\kappa\over 2\pi|\kappa|}
\end{equation}
for $N \ge 3$.  It is interesting to see that for $SO(3)$, which is also the adjoint representation of $SU(2)$, there is no correction in the Higgs phase.  For $SO(2)$, the gauge symmetry is again completely broken in the Higgs phase and the correction is also vanishing.  This is consistent with the results found in Ref. \cite{CHTh}.  Thus, we see that the abelian result is really just a special case of the non-abelian ones.

Naturally, it is interesting to see whether this kind of analysis can be applied to the case that the Higgs field is in the adjoint representation.  Since there can be several inequivalent vacua in these systems, it is non-trivial to show that the Chern-Simons coefficient is quantized in all the Higgs phases.  At this moment, there are at least two difficulties.  First, we do not know the form of the projection operators such as $P_1, P_2$ and $P_3$ in the adjoint representation.  Second, there could be an infinite number of would be Chern-Simons terms, e.g. $i\epsilon^{\mu\nu\rho} \tr\bigl\{[(\Phi^\dagger)^n, F_{\nu\rho}] (D_\mu \Phi^n) -(D_\mu\Phi^n)^\dagger [F_{\nu\rho}, \Phi^n]\bigr\},$ with $n$ an arbitrary positive integer.  As mentioned above, this may also make it impossible for us to find the correction to the Chern-Simons term by calculating only the parity odd part of the vacuum polarization in the unbroken sector.

\vskip1cm
\noindent{\bf Acknowledgements}

The work of H.-C. K. is supported in part by the National Science Council of R.O.C. under grant No. NSC87-2112-M-032-002.

\newpage
\appendix
\centerline{\bf Appendix}
\begin{eqnarray}
\Pi^{Ic}(p) = \quad
&\;& \int {d^3k\over (2\pi)^3}
\biggl\{ {M\bigl[k^2 p^2 -(k\cdot p)^2 \bigr] \over
p^2 (k^2+M_{W^+}^2)(k^2+M_{W^-}^2)(k-p)^2 \bigl[(k-p)^2+ M^2\bigr] }\biggr\}\nonumber\\
&\;& \qquad\quad \times \biggl\{ \bigl[5M^2 + 8k^2 - 12k\cdot p + 8p^2 \bigr] \nonumber\\
&\;&\qquad\qquad + {(k^2 + M_{W^+} M_{W^-})
\bigl[- M^2 + 2k^2 + 2k\cdot p \bigr] \over k^2} \biggr\} \nonumber\\
+ &\;& \int {d^3k\over (2\pi)^3} 
\biggl\{ {-2M(k\cdot p)\bigl[k^2 - p^2 \bigr] \over
p^2 (k^2 + M_{W^+}^2)(k^2+M_{W^-}^2)(k-p)^2} \nonumber\\
&\;& \qquad\qquad + {M(k^2 + M_{W^+} M_{W^-}) 
\bigl[-k^2p^2 - (k\cdot p)^2 + 2k^2(k\cdot p) \bigr]
\over p^2 k^2 (k^2 + M_{W^+}^2)(k^2+M_{W^-}^2)(k-p)^2} \nonumber\\
&\;& \qquad\qquad + { M M_{W^+} M_{W^-} \bigl[k\cdot p - p^2\bigr]
+ M\bigl[-k^2p^2 - (k\cdot p)^2 + 2p^2(k\cdot p)\bigr]
\over p^2 k^2 (k-p)^2 \bigl[(k-p)^2 + M^2\bigr]} \biggr\}, \nonumber\\
\Pi^{Id}(p) = \quad
&\;& \int {d^3k\over (2\pi)^3}
\biggl\{ {M\bigl[k^2 p^2 -(k\cdot p)^2 \bigr] \over
p^2 (k^2+M_{W^+}^2)(k^2+M_{W^-}^2)
\bigl[(k-p)^2+ M_{Z^+}^2\bigr]\bigl[(k-p)^2 +M_{Z^-}^2\bigr] }\biggr\}\nonumber\\
&\;& \qquad\quad \times \biggl\{ 6M^2 
+ {(k^2 + M_{W^+} M_{W^-})
\bigl[- M^2 + 8k^2 - 4k\cdot p + 4p^2 \bigr] \over k^2} \\
&\;& \qquad\qquad + {\bigl[(k-p)^2 + M_{Z^+} M_{Z^-}\bigr] 
\bigl[-M^2 + 8k^2 - 12k\cdot p + 8p^2 \bigr] \over (k-p)^2 } \nonumber\\
&\;& \qquad\qquad + {(k^2 + M_{W^+} M_{W^-})\bigl[(k-p)^2 + M_{Z^+} M_{Z^-} \bigr]
\bigl[-6k^2 + 6k\cdot p -4p^2 \bigr] \over k^2(k-p)^2 }
\biggr\} \nonumber\\
+ &\;& \int {d^3k\over (2\pi)^3} 
\biggl\{ {-M(k\cdot p)\bigl[M_{W^+} M_{W^-} + 2k^2 - 2p^2 \bigr] \over
p^2 (k^2 + M_{Z^+}^2)(k^2+M_{Z^-}^2)(k-p)^2} \nonumber\\
&\;&\qquad\qquad +  {-M(k\cdot p)\bigl[M_{Z^+} M_{Z^-} + 2k^2 - 2p^2 \bigr] \over
p^2 (k^2 + M_{W^+}^2)(k^2+M_{W^-}^2)(k-p)^2} \nonumber\\
&\;& \qquad\qquad + {M(k^2 + M_{W^+} M_{W^-}) 
\bigl[-k^2p^2 - (k\cdot p)^2 + 2k^2(k\cdot p)\bigr]
\over p^2 k^2(k^2 + M_{W^+}^2)(k^2+M_{W^-}^2)(k-p)^2} \nonumber\\
&\;& \qquad\qquad + {M(k^2 + M_{Z^+} M_{Z^-}) 
\bigl[-k^2p^2 - (k\cdot p)^2 + 2k^2(k\cdot p)\bigr]
\over p^2 k^2(k^2 + M_{Z^+}^2)(k^2+M_{Z^-}^2)(k-p)^2} \biggr\}. \nonumber
\end{eqnarray}
In each of the above expressions, the first integral is identical to that of the corresponding Feynmann diagram in the usual Landau gauge, and the second integral comes from the combining effect of the ghost and unphysical Higgs.
\begin{eqnarray}
\Pi^{IIa}(p) = &\;& \int {d^3k\over (2\pi)^3} \biggl\{ 
{M(g^2\vp^2) (k\cdot p) \over 
p^2 (k^2+M_{W^+}^2)(k^2+M_{W^-}^2)\bigl[(k-p)^2+ m_H^2\bigr] }\biggr\}\nonumber\\
\Pi^{IIb}(p) = &\;& \int {d^3k\over (2\pi)^3} \biggl\{ 
{M(g^2\vp^2) (k\cdot p) \over 
p^2 (k^2+M_{W^+}^2)(k^2+M_{W^-}^2)(k-p)^2 }\biggr\}\nonumber\\
\Pi^{IIc}(p) = &\;& \int {d^3k\over (2\pi)^3} \biggl\{ 
{M(g^2\vp^2) (k\cdot p) \over 
p^2 k^2(k^2+M^2)(k-p)^2 }\biggr\}\nonumber\\
\Pi^{IId}(p) = &\;& \int {d^3k\over (2\pi)^3} \biggl\{ 
{M(g^2\vp^2) (k\cdot p) \over 
p^2 (k^2+ M_{Z^+}^2)(k^2+ M_{Z^-}^2)(k-p)^2 }\biggr\}\\
\Pi^{IIe}(p) = &\;& \int {d^3k\over (2\pi)^3} \biggl\{ 
{M(g^2\vp^2) (k\cdot p) \over 
p^2 (k^2+ M_{Z^+}^2)(k^2+ M_{Z^-}^2)\bigl[(k-p)^2+ m_H^2\bigr] }\biggr\}\nonumber\\
\Pi^{IIf}(p) = &\;& \int {d^3k\over (2\pi)^3} \biggl\{ 
{M(g^2\vp^2) (k\cdot p) \over 
p^2 (k^2+M_{W^+}^2)(k^2+M_{W^-}^2)(k-p)^2 }\biggr\}. \nonumber 
\end{eqnarray}

\vfill\eject

\vspace*{-8.5cm}
\begin{picture}(18000,30000)
\drawloop\gluon[\NE 0](7500,0)
\drawarrow[\W\ATBASE](8000,2420)
\drawarrow[\E\ATBASE](6800,-2450)
\put(6800, 2700){$k-p$}
\put(6400, -3400){$k$}
\put(4000,2000){$Q^{m'}_\rho$}
\put(3500,-2500){$Q^{n'}_{\rho^\prime}$}
\put(9700,2000){$Q^{m'}_\lambda$}
\put(9700,-2500){$Q^{n'}_{\lambda^\prime}$}

\frontstemmed\drawline\gluon[\E\CENTRAL](\loopbackx,\loopbacky)[6]
\global\advance\pmidx by -100
\global\advance\pmidy by 600
\drawarrow[\E\ATBASE](\pmidx,\pmidy)
\global\advance\pmidy by 700
\put(\pmidx,\pmidy){$p$}
\global\advance\gluonbackx by -1000
\global\advance\gluonbacky by 1200
\put(\gluonbackx,\gluonbacky){$A^n_\nu$}

\frontstemmed\drawline\gluon[\W\FLIPPEDCENTRAL](\loopfrontx,\loopfronty)[6]
\global\advance\pmidx by -100
\global\advance\pmidy by 600
\drawarrow[\E\ATBASE](\pmidx,\pmidy)
\global\advance\pmidy by 700
\put(\pmidx,\pmidy){$p$}
\global\advance\gluonbacky by 1200
\put(\gluonbackx,\gluonbacky){$A^m_\mu$}
\put(6500,-6000){\small{(a)}}

\drawline\gluon[\E\CENTRAL](20000,0)[16]
\global\advance\pmidy by 600
\put(\pmidx,\pmidy){\oval(5000,5000)[t]}
\global\advance\pmidx by -200
\drawarrow[\E\ATBASE](\pmidx,\pmidy)
\global\advance\pmidx by -200
\global\advance\pmidy by -1700
\put(\pmidx,\pmidy){$k$}
\global\advance\pmidx by 200
\global\advance\pmidy by 4250
\drawarrow[\W\ATTIP](\pmidx,\pmidy)
\global\advance\pmidx by -1200
\global\advance\pmidy by 400
\put(\pmidx,\pmidy){$k-p$}
\put(23200,3000){$\phi_{b,q}$}
\put(24500,-1500){$Q^{m'}_\mu$}
\put(29500,3000){$\phi_{a,p}$}
\put(29500,-1500){$Q^{n'}_\nu$}

\global\advance\pfronty by 1200
\put(\pfrontx,\pfronty){$A^m_\mu$}
\global\advance\pfrontx by 2200
\global\advance\pfronty by -600
\drawarrow[\E\ATBASE](\pfrontx,\pfronty)
\global\advance\pfronty by 700
\put(\pfrontx,\pfronty){$p$}

\global\advance\pbackx by -1000
\global\advance\pbacky by 1200
\put(\pbackx,\pbacky){$A^n_\nu$}
\global\advance\pbackx by -1500
\global\advance\pbacky by -600
\drawarrow[\E\ATBASE](\pbackx,\pbacky)
\global\advance\pbacky by 700
\put(\pbackx,\pbacky){$p$}

\put(26500,-6000){\small{(b)}}
\put(-3000,-8000){\small{
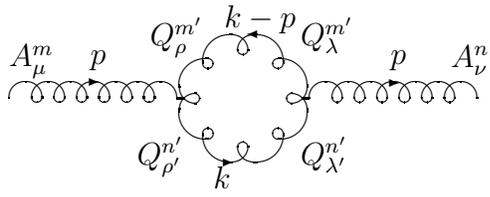
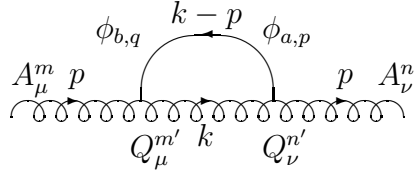
Fig.1. The one-loop diagrams that contribute to the parity odd part of the 
vacuum }}
\put(-3000,-9500){\small{
polarization.  Fig. 1(a) involves an internal gluon loop, while Fig. 1(b) involves an}}
\put(-3000,-11000){\small{
internal loop with both gluon and Higgs field.}}
\end{picture}
\vfill\eject

\end{document}